\documentstyle[12pt]{article}

\makeatletter
\@addtoreset{equation}{section}
\makeatother

\begin{document}
\begin{center}
{\bf A STATISTICAL SUPERFIELD AND ITS OBSERVABLE CONSEQUENCES} \\[.5in]

Roland E. Allen \\[.25in]

{\it Center for Theoretical Physics, Texas A\&M University, \\
College Station, Texas  77843, USA}
\end{center}
\bigskip
\bigskip
\begin{abstract}
A new kind of fundamental superfield is proposed, with an Ising-like
 Euclidean action.  Near the Planck energy it undergoes its first stage
of symmetry-breaking, and the ordered phase is assumed to support specific
kinds of topological defects.  This picture leads to a low-energy Lagrangian
 which is similar to that of standard physics, but there are interesting and
 observable differences.  For example, the cosmological constant vanishes,
 fermions have an extra coupling to gravity, the gravitational interaction 
of W-bosons is modified, and Higgs bosons have an unconventional equation of 
motion.
\end{abstract}
\bigskip
\bigskip
\bigskip
\bigskip
\bigskip
\bigskip
\bigskip
\bigskip
\bigskip
\bigskip
e-mail:  allen@phys.tamu.edu \\
tel.:  (409) 845-4341 \\
fax:  (409) 845-2590
\bigskip
\bigskip
\bigskip
\bigskip
\bigskip
\bigskip
\begin{center}
{\bf International Journal of Modern Physics A, Vol. 12, No. 13 (1997) 2385-2412}
\end{center}
\bigskip
\bigskip
\bigskip
CTP-TAMU-15/96
\pagebreak
\section{Introduction}

The terms ``superfield'' and ``supersymmetry'' are ordinarily used in a context
which presupposes local Lorentz invariance.$^{1-3}$  It is far from clear, 
however,
that Lorentz invariance is still valid near the Planck scale, fifteen orders
of magnitude above the energies where it has been tested.  (A century ago, all
accepted theories presupposed Galilean invariance.)  In this paper the above
terms will be used in a broader sense, to mean any field which has both
commuting and Grassmann parts and any symmetry which relates these parts.  At
the same time, the presumption of Lorentz invariance at arbitrarily high
energies will be replaced by a less stringent requirement: Lorentz invariance,
and the other principles of standard physics, need only emerge at the relatively
low energies where they have been tested.  One is then free to consider any
description which is mathematically consistent and also consistent with
experiment and observation.

It appears, however, that a fundamental theory still needs four central
ingredients: a space, a field, an action, and a pattern of symmetry-breaking. 
The specific versions assumed here are as follows:

\indent
(1)  The space (or base manifold) is $R^D$; i.e., it is $D$-dimensional, flat,
and initially Euclidean.

\indent
(2)  The classical field $\Psi$ at each point $x$ is an $N$-dimensional
supersymmetric vector.

\indent
(3)  The Euclidean action $S$ has the same basic form as the Hamiltonian for
spins on a lattice.$^4$

\indent
(4)  Below the Planck energy $<\Psi>$ becomes nonzero.  This order parameter
can then support topological defects, analogous to those in condensed matter
physics.$^{5-11}$  Three such defects are postulated in Section 7.  The first
two of these cause the original symmetry group $G_0$ to break down locally to a
reduced symmetry group U(1) x SU(2) x G.  They also produce a local
``filamentary'' geometry with 4 extended dimensions and ($D$-4) that are compact.

After a reasonable series of approximations, we will find that this relatively
simple picture leads to a low-energy Lagrangian (9.4) which closely resembles
the conventional Lagrangian of particle physics$^{12,13}$ and general
relativity.$^{14,15}$  There are some interesting and observable differences,
however, and these are discussed at the end of the paper.

\pagebreak
\noindent
\section{The superfield and its Euclidean action}

First consider an analogy from ordinary statistical mechanics: classical spins
on a lattice in $D$ dimensions.  Each spin $s(x)$ is an $N$-dimensional vector
whose components are real numbers.  In the continuum limit, one can obtain a
Ginzburg-Landau Hamiltonian

\begin{equation}
H = \int d^D x \left[ \frac{1}{2m} \partial^M s^{\dagger} \partial_M s - \mu
s^{\dagger} s + \frac{1}{2} b \left( s^{\dagger} s \right) ^2 \right] 
\end{equation}

\noindent
where $\partial_M~=~\partial/\partial x^M$.  (See, e.g., (3.5.17) of Ref. 4. 
Summation is implied over repeated indices, and inner products involving
vectors are also implied.)  The physical properties of this statistical system
are determined by $H$, via the partition function.$^{4,16-19}$  At low
temperature the order parameter $<s>$ can become nonzero, making the system
ferromagnetic.

The starting point of the present theory is very similar: a classical field
$\Psi(x)$ having the form

\begin{equation}
\Psi = \left( \begin{array}{c}
                 z_{1} \\ z_{2} \\ \vdots \\ z_{N}
              \end{array} \right)  \end{equation}

\noindent
where each $z$ consists of an ordinary complex number $z_b$ and a complex
Grassmann number $z_f$:

\begin{equation}
z = \left( \begin{array}{c}
              z_{b} \\ z_{f}
           \end{array} \right) .  \end{equation}

\noindent
(Anticommuting Grassmann numbers are required in any classical treatment which
includes fermions.$^{3,4,12,18-26}$)  The Euclidean action is postulated to
have the Ising-like form

\begin{equation}
S~=~\beta H  
\end{equation}

\begin{equation}
H~=~-J \sum_{ij} \Psi_i^{\dagger} \Psi_j + {1 \over 2} r\sum_{ij} \left(
\Psi_i^{\dagger} \Psi_j \right) ^2  
\end{equation}

\noindent
where the summation is over nearest-neighbor lattice sites.  The first term
represents a tendency for the field to be aligned at neighboring points, and
the second ensures that $S$ has a lower bound.

Since

\begin{equation}
\Psi_i^{\dagger} \Psi_j + \Psi_j^{\dagger} \Psi_i = -\left( \Psi_i - \Psi_j
\right)^{\dagger} \left( \Psi_i - \Psi_j \right) + \Psi_i^{\dagger} \Psi_i +
\Psi_j^{\dagger} \Psi_j 
\end{equation}

\noindent
the continuum version of (2.4) is

\begin{equation}
S = \int d^D x \left[ \frac{1}{2m} \partial^M \Psi^{\dagger} \partial_M \Psi -
\mu \Psi^{\dagger} \Psi + \frac{1}{2} b \left( \Psi^{\dagger} \Psi \right) ^2
\right]  
\end{equation}

\noindent
where $\Psi(x)$ = $(a^{-D} \beta)^{1/2} \Psi_j$, $(2m)^{-1}$ = $a^2J$, $\mu$ =
2$DJ$, $b$ = 2$Da^D \beta^{-1}r$, and $a$ is the lattice spacing.  We
will find below, in (7.28) and (7.34), that $m$, $\mu$, and $b$ can be related to
the Planck energy $m_P$, defined by

\begin{equation}
m_P^{-1}~=~\ell_P~=~(16\pi G)^{1/2} 
\end{equation}

\noindent
where $G$ is the gravitational constant.  (Units with $\hbar$ = c = 1 are used,
so mass and energy are equivalent to inverse length.)  The definition (2.8)
implies that $\ell_P~\sim~10^{-32}$ cm and $m_P~\sim~10^{15}$ TeV.

In the continuum treatment represented by (2.7), the partition function becomes
a Euclidean path integral.$^{4,19,21,24,25,27,28}$  It initially has the form

\begin{equation}
Z~=~N_1 \int {\cal D}(Re \Psi ) {\cal D} (Im \Psi )e^{-S} 
\end{equation}

\noindent
but can be rewritten in the equivalent form

\begin{equation}
Z~=~N_2 \int {\cal D} \Psi {\cal D} \Psi^{\dagger} e^{-S}  
\end{equation}

\noindent
where $N_1$ and $N_2$ are constants.  In (2.10), and in the following, the
functions $\Psi$ and $\Psi^{\dagger}$ are taken to vary independently.$^{29}$

$S$ can be interpreted as the Euclidean action for interacting Bose and Fermi
fields $\Psi_b$ and $\Psi_f$:

\begin{equation}
S~=~S_b^{(0)} + S_f^{(0)} + S_{int} 
\end{equation}

\noindent
with

\begin{eqnarray}
S_b^{(0)}~&=&~ \int d^D x \left( \frac{1}{2m} \partial^M \Psi_b^{\dagger}
\partial_M \Psi_b - \mu \Psi_b^{\dagger}
\Psi_b \right)  \\
S_f^{(0)}~&=&~ \int d^D x \left( \frac{1}{2m} \partial^M \Psi_f^{\dagger}
\partial_M \Psi_f - \mu \Psi_f^{\dagger} \Psi_f \right) \\
S_{int}~&=&~ \int d^D x~{1 \over 2} b \left( \Psi_b^{\dagger} \Psi_b +
\Psi_f^{\dagger} \Psi_f \right) ^2 . 
\end{eqnarray}

\noindent
Notice that $S$ is supersymmetric in an unconventional way: $\Psi_b$ and $\Psi_f$
have the same number of components and the same form.  There is no
contradiction with the spin-and-statistics theorem$^{23}$ because this theorem
is based on Lorentz invariance, a symmetry that will emerge only at low energies,
and after a Wick rotation to Lorentzian time.

Although Lorentz transformations can be defined only at a later stage, $S$ is
already invariant under general coordinate transformations.  To make this
explicit, we should replace $d^D x$ by the invariant volume element $d^D x~h$,
where

\begin{equation}
h~=~\left( det~h_{MN} \right)^{1/2}
\end{equation}

\noindent
and $h_{MN}$ is the metric tensor for flat Euclidean space, initially given by

\begin{equation}
h_{MN}~=~\delta_{MN} . 
\end{equation}

\noindent
$\Psi_b$ and $\Psi_f$ are both taken to transform as scalars.  This is
consistent with the usual convention in general relativity, according to which
a spinor transforms as a scalar under general coordinate
transformations.$^{15}$  After the Lagrangian of (9.4) has been obtained, we
will additionally have Lorentz transformations,$^{14,15}$ with the usual behavior
for spinors and the usual connection between spin and statistics.

\pagebreak
\section{The order parameter}

$S$ has the same form as the grand-canonical Hamiltonian for a conventional
superfluid.$^{5}$  This Ginzburg-Landau form indicates that $<\Psi_b>$ will
be nonzero at low temperature, so it is natural to write

\begin{equation}
\Psi_b~=~ \Psi_s + \Phi_b 
\end{equation}

\noindent
as in Ref. 5.  The classical equations of motion for the order parameter
$\Psi_s$, the bosonic excitations $\Phi_b$, and the fermionic excitations
$\Psi_f$ follow from

\begin{equation}
\delta S~=~0 
\end{equation}

\noindent
with $\Psi_b$, $\Psi_b^{\dagger}$, $\Psi_f$, and $\Psi_f^{\dagger}$ all varied
independently.

We will consider fermionic excitations in the next section and bosonic
excitations in Section 8.  For the moment, however, let us focus on the order
parameter.  After integration by parts (with boundary terms assumed to vanish)
(2.11) becomes

\begin{equation}
S_0~=~ \int d^D x~h \Psi_s^{\dagger} \left( T + \frac{1}{2} V - \mu \right)
\Psi_s 
\end{equation}

\noindent
in the absence of excitations, where

\begin{equation}
T~=~-\frac{1}{2m} \partial^M \partial_M 
\end{equation}

\begin{equation}
V(x)~=~bn_s 
\end{equation}

\begin{equation}
n_s~=~ \Psi_s^{\dagger} \Psi_s~~~. 
\end{equation}

\noindent
Then (3.2) gives

\begin{equation}
(T + V - \mu) \Psi_s = 0 
\end{equation}

\noindent
and

\begin{equation}
\Psi_s^{\dagger} (T + V - \mu) = 0 
\end{equation}

\noindent
with the operator acting to the left in this last equation.

For an ordinary superfluid like $^4$He, one writes
$\Psi_s~=~n_s^{1/2}~exp(i\theta$).  The appropriate generalization is

\begin{equation}
\Psi_s~=~n_s^{1/2} U \eta 
\end{equation}

\begin{equation}
\Psi_s^{\dagger}~=~\eta^{\dagger} U^{\dagger} n_s^{1/2} 
\end{equation}

\noindent
with

\begin{equation}
U^{\dagger} U~=~1 
\end{equation}

\begin{equation}
\eta^{\dagger} \eta~=~1 . 
\end{equation}

\noindent
$U(x)$ and $U^{\dagger}(x)$ are matrices, and $\eta$ and $\eta^{\dagger}$ are
constant vectors.  (Recall that $\Psi$ and $\Psi^{\dagger}$ vary independently,
so the quantities in (3.10) are not necessarily the Hermitian conjugates of
those in (3.9).)  For $^4$He, the superfluid velocity is defined by
$m\vec{v} = \vec{\nabla} \theta$.  The
generalization is

\begin{equation}
mv_M~=~ -iU^{-1} \partial_M U. 
\end{equation}

\noindent
Notice that (3.11) gives $\partial_M U^{\dagger} U = -U^{\dagger} \partial_M U$
with $U^{\dagger} = U^{-1}$, or

\begin{equation}
mv_M~=~i \partial_M U^{\dagger} U . 
\end{equation}

When (3.9) -- (3.14) are used in (3.7), the result is

\begin{equation}
\eta^{\dagger} n_s^{1/2} \left[ \left( {1 \over 2} mv^M v_M + V - {1 \over {2m}}
\partial^M \partial_M - \mu \right) -i \left( {1 \over 2} \partial^M v_M + v^M
\partial_M \right) \right] n_s^{1/2} \eta = 0. 
\end{equation}

\noindent
The Schr\"{o}dinger-like equations (3.7) and (3.8) also lead to the equation of
continuity

\begin{equation}
\partial^M j_M~=~0 
\end{equation}

\noindent
with

\begin{equation}
j_M~=~ \frac{1}{2im} \left[ \Psi_s^{\dagger} \left( \partial_M \Psi_s \right) -
\left( \partial_M \Psi_s^{\dagger} \right) \Psi_s \right] 
\end{equation}

\begin{equation}
= \eta^{\dagger} n_s v_M \eta . 
\end{equation}

\noindent
Substitution of (3.18) into (3.16) gives

\begin{equation}
\eta^{\dagger} \left( \partial^M v_M + v^M \partial_M \right) n_s \eta =0
\end{equation}

\noindent
reducing (3.15) to

\begin{equation}
\frac {1}{2} m\bar{v} ^2 + V + P = \mu 
\end{equation}

\noindent
where

\begin{equation}
\bar{v}^2 = \eta^{\dagger} v^M v_M \eta 
\end{equation}

\begin{equation}
P~=~- \frac{1}{2m} n_s^{-1/2} \partial^M \partial_M n_s^{1/2} . 
\end{equation}

\noindent
Eq. (3.20) is a quantum version of Bernoulli's equation, with part of the kinetic
energy playing the role of pressure.

In the next section we will consider an order parameter with the symmetry group
U(1) x SU(2).  For this case, the ``superfluid velocity'' $v_M$ can be written
in terms of the identity matrix $\sigma_0$ and the Pauli matrices $\sigma_a$ :

\begin{equation}
v_M~=~v_M^{\alpha} \sigma_{\alpha} ,~~~\alpha ~=~0,1,2,3. 
\end{equation}

\noindent
It is reasonable to assume that the order parameter has the symmetry

\begin{equation}
\eta^{\dagger} \sigma_a \eta ~=~0 
\end{equation}

\noindent
which means that the system is not spin-polarized when $v_M$ = 0.  Then cross
terms involving $\sigma_a \sigma_0$ vanish, and the relation

\begin{equation}
\sigma_a \sigma_b + \sigma_b \sigma_a = 2 \delta_{ab} 
\end{equation}

\noindent
further reduces (3.20) to

\begin{equation}
{1 \over 2} mv_{\alpha}^M v_M^{\alpha} + V + P = \mu . 
\end{equation}

\pagebreak
\section{Fermionic excitations, U(1) x SU(2) order parameter}

When fermionic excitations are included, (2.11) becomes $S = S_0 + S_f$, with

\begin{equation}
S_f ~=~ \int d^D x~h  \Psi_f^{\dagger} \left( T + V - \mu \right) \Psi_f .
\end{equation}

\noindent
The term involving $\left( \Psi_f^{\dagger} \Psi_f \right) ^2$ is neglected in
comparison to the one containing $V$ because fermions cannot form a condensate.

According to (3.2), $\Psi_f$ obeys the same equation of motion as $\Psi_s$, and
will share its rapid oscillations in regions where $\mu$--$V$ is large.  In order
to eliminate these oscillations, it is convenient to write

\begin{equation}
\Psi_f ~=~U\psi_f~=~n_s^{1/2} U \stackrel{\sim}{\psi}_f . 
\end{equation}

\noindent
For simplicity, $n_s$ will initially be regarded as slowly varying.  Then we
will find that low-energy, long-wavelength excitations $\psi_f$ also correspond
to low values of the action (4.1), and that it is consistent to identify $\psi_f$
with the fermionic fields observed in nature.

First consider the case $N$=2, $D$=4, with symmetry group U(1) x SU(2) for the
order parameter.  The coordinates are

\begin{equation}
x^{\mu} ,~~~\mu~=~0,1,2,3. 
\end{equation}

\noindent
Substitution of (4.2) into (4.1) gives

\begin{equation}
S_f~=~ \int d^4 x~h \psi_f^{\dagger} \left[ \left( \frac{1}{2} mv^{\mu} v_{\mu}
+ V - \frac{1}{2m} \partial^{\mu} \partial_{\mu} - \mu \right) - i \left(
\frac{1}{2} \partial^{\mu} v_{\mu} + v^{\mu} \partial_{\mu} \right) \right]
\psi_f . 
\end{equation}

\noindent
For low-energy (long-wavelength) excitations, $\partial^{\mu} \partial_{\mu}
\psi_f$ can be neglected in comparison with $mv^{\mu} \partial_{\mu} \psi_f$.  If
$n_s$ is slowly varying, $P$ can also be neglected.  Then the Bernoulli equation
(3.26), together with (3.23) and (3.25), implies that

\begin{equation}
\frac{1}{2} mv^{\mu} v_{\mu} + V - \mu = mv_0^{\mu} v_{\mu}^a \sigma_a .
\end{equation}

In Section 7, a cosmological picture will be presented in which $v_{\mu}^a$ is
real but $v_{\mu}^0$ is imaginary,

\begin{equation}
\frac{1}{2} mv_0^{\mu} v_{\mu}^0 < 0, 
\end{equation}

\noindent
and the basic texture is given by

\begin{equation}
v_k^0 ~=~ v_0^a = 0~for~k,a = 1,2,3. 
\end{equation}

\noindent
Then (4.4) becomes

\begin{equation}
S_f ~=~\int d^4 x~h \psi_f^{\dagger} \left( - \frac{1}{2} i \partial^{\mu} v_{\mu}
- iv^{\mu} \partial_{\mu} \right) \psi_f 
\end{equation}

\noindent
or

\begin{equation}
S_f ~=~\int d^4 x~h {1 \over 2} \left[ \psi_f^{\dagger} v^{\mu} \left( - i
\partial_{\mu} \psi_f \right) + \left( - i \partial_{\mu} \psi_f \right)
^{\dagger} v^{\mu} \psi_f \right] 
\end{equation}

\noindent
after integration by parts.  There is no reason why the texture of (4.7) must
be perfectly rigid, however, so we should permit small deformations $v_k^0$ and
$v_0^a$.  When second-order terms are neglected, (4.9) is changed to

\begin{equation}
S_f~=~ \int d^4 x~\bar{\cal L}_f 
\end{equation}

\begin{equation}
\bar{\cal L}_f~=~-{1 \over 2} ih\psi_f^{\dagger} v_{\alpha}^{\mu}
\sigma^{\alpha} \stackrel{\sim}{\nabla}_{\mu} \psi_f + conj. 
\end{equation}

\noindent
where

\begin{equation}
\stackrel{\sim}{\nabla}_{\mu}~=~ \partial_{\mu} + \Gamma_{\mu} +ia_{\mu}
+ib_{\mu} 
\end{equation}

\noindent
with

\begin{equation}
a_0~=~0,~~~a_k = mv_k^0~~~~(k = 1,2,3) 
\end{equation}

\begin{equation}
b_k~=~0,~~~b_0 = mv_0^a \sigma_a~~~~(a = 1,2,3). 
\end{equation}

\noindent
Here ``conj.'' represents a second term like that in (4.9).  After the
transformation to Lorentzian time in Section 9, it can be regarded as the true
Hermitian conjugate of the first term, represented by ``h.c.''  The spin
connection $\Gamma_{\mu}$ is initially zero, but must be added to (4.11) to
compensate for local transformations of $\psi_f$ when frame rotations are
permitted.$^{15}$

Suppose that we now define an effective vierbein $e_{\mu}^{\alpha}$ and an
effective metric tensor $g_{\mu \nu}$ by

\begin{equation}
e_{\alpha}^{\mu} = v_{\alpha}^{\mu},~~~\mu , \alpha = 0,1,2,3 
\end{equation}

\begin{equation}
e_{\mu}^{\alpha} e_{\beta}^{\mu} = \delta_{\beta}^{\alpha} 
\end{equation}

\begin{equation}
g_{\mu \nu} = \eta_{\alpha \beta} e_{\mu}^{\alpha} e_{\nu}^{\beta} . 
\end{equation}

\noindent
The Minkowski metric tensor $\eta_{\alpha \beta}$ = diag (-1,1,1,1) is needed
because of (4.6) and the requirement that a Euclidean metric tensor $g_{\mu
\nu}$ should have signature (++++).  $\bar{\cal L}_f$ then has nearly
the same form as the standard Euclidean Lagrangian for massless spin 1/2
fermions in the Weyl representation.$^{13,19,25,30-32}$  There are two
differences: the extra couplings $a_{\mu}$ and $b_{\mu}$, and the factor of $h$
rather than

\begin{equation}
g~=~ \left( det~g_{\mu \nu} \right) ^{1/2} . 
\end{equation}

\noindent
These features will be discussed near the end of the paper, but suppose that we
momentarily disregard them.  The behavior of massless fermions will then be the
same as in a curved spacetime with metric tensor $g_{\mu \nu}$.  In the
present theory, the geometry of spacetime is defined by the texture of the
order parameter, with the ``superfluid velocity'' $v_{\alpha}^{\mu}$ becoming
the vierbein $e_{\alpha}^{\mu}$.  The origin of spacetime curvature will be
discussed in Section 7, and the transformation to Lorentzian time in Section 9.

\pagebreak
\section{Fermionic excitations, U(1) x SU(2) x G order parameter}

The treatment of the preceding section contains only one fermion species and no
forces other than gravity.  Let us now move to a more realistic description,
with $N>$2 and $D>$4, which is similar to standard higher-dimensional
theories.$^{3,33-35}$  The ordered phase described by $\Psi_s$ is assumed to
locally have a ``filamentary'' geometry, with 4 extended dimensions and
{\it d} that are compact.  To be more precise, it occupies only a very
small volume

\begin{equation}
V_B~=~ \int d^d x 
\end{equation}

\noindent
in an internal space $x_B$ with coordinates

\begin{equation}
x^m,~~~m~=~4,5,\ldots,3 + d \nonumber
\end{equation}

\noindent
but a very large volume in the 4-dimensional external spacetime $x_A$ with
coordinates $x^{\mu}$.  In the simplest picture, the ordered phase is a
$d$-dimensional ball of condensate in internal space, which can be described by
one radial coordinate and ($d$--1) angular coordinates.  To avoid confusion,
however, we will retain the original rectangular coordinates $x^m$ in the
discussion below, so that det $h_{mn}$ = 1 and $h$ = (det $h_{MN})^{1/2}$ = (det
$h_{\mu \nu})^{1/2}$.

It is also assumed that the order parameter locally has the form of a 
product:

\begin{equation}
\Psi_s~=~\Psi_A \Psi_B 
\end{equation}

\noindent
where $\Psi_A$ has the symmetry group U(1) x SU(2) and $\Psi_B$ has an
unspecified symmetry group G with generators $\sigma_c$.  Then (3.23) must be
generalized to

\begin{equation}
v_M~=~v_M^A \sigma_A = v_M^{\alpha} \sigma_{\alpha} + v_M^c \sigma_c,~~~c \geq
4. 
\end{equation}

Both the filamentary geometry and the form of the order parameter originate
from two topological defects discussed in Section 7, associated respectively
with the symmetry groups G and U(1) x SU(2).

In generalizing the definition of the effective vierbein $e_{\mu}^{\alpha}$, it
will be convenient to choose

\begin{equation}
e_M^c~=~v_M^c 
\end{equation}

\noindent
while retaining (4.15) and (4.16):

\begin{equation}
e_{\alpha}^M~=~v_{\alpha}^M 
\end{equation}

\begin{equation}
e_M^{\alpha} e_{\beta}^M = \delta_{\beta}^{\alpha} . 
\end{equation}

\noindent
The effective metric tensor is then

\begin{equation}
g_{MN}~=~\eta_{AB} e_M^A e_N^B = \eta_{\alpha \beta} e_M^{\alpha} e_N^{\beta} +
e_M^c e_N^c. 
\end{equation}

Standard Kaluza-Klein theory$^{33,34}$ begins with an unperturbed metric tensor
having the form $g_{\mu \nu}$ = $g_{\mu \nu} \left( x_A \right)$,
$g_{mn}$ = $g_{mn} \left( x_B \right)$, $g_{\mu m}$ = $g_{m \mu}$ = 0.  The
present theory similarly begins with an unperturbed order parameter having the
form

\begin{equation}
\Psi_s = \Psi_A \left( x_A \right) \Psi_B \left( x_B \right) 
\end{equation}

\noindent
which implies the texture

\begin{equation}
v_{\alpha}^{\mu} = v_{\alpha}^{\mu} \left( x_A \right) 
\end{equation}

\begin{equation}
v_{\mu}^c = 0 
\end{equation}

\begin{equation}
v_m^c = v_m^c \left( x_B \right) 
\end{equation}

\begin{equation}
v_{\alpha}^m = 0 
\end{equation}

\noindent
and the effective geometry $g_{\mu \nu}$ = $g_{\mu \nu} \left( x_A
\right), g_{mn} = g_{mn} \left( x_B \right)$, $g_{\mu m} =
g_{m \mu} = 0$.

The form (5.9) requires that $n_A = \Psi_A^{\dagger} \Psi_A$ and $\mu_A =
\frac{1}{2} mv_{\alpha}^{\mu} v_{\mu}^{\alpha}$ be regarded as constant in
treating the rapid variations of the internal order parameter $\Psi_B$.  Then
(3.7) gives

\begin{equation}
\left( - {1 \over {2m}} \partial^m \partial_m + V \right) \Psi_B = \mu_B \Psi_B
\end{equation}

\noindent
where $V\left( x_B \right)$ = $bn_A n_B \left( x_B \right), n_B =
\Psi_B^{\dagger} \Psi_B$, and $\mu_A = \mu - \mu_B$.  The internal versions of
(3.9) and (3.13) are

\begin{equation}
\Psi_B = n_B^{1/2} U_B \eta_B 
\end{equation}

\noindent
and

\begin{equation}
mv_m = -iU_B^{-1} \partial_m U_B 
\end{equation}

\noindent
with

\begin{equation}
v_m = v_m^c \sigma_c . 
\end{equation}

Now let us turn to the fermion field $\psi_f$ of (4.2).  It can be expanded in a
complete set of states $\psi_r^B \left( x_B \right)$ with coefficients $\psi_r
\left(x_A \right):^{3,33-35}$

\begin{equation}
\psi_f \left( x_A , x_B \right) = \sum_r \psi_r \left( x_A \right) \psi_r^B
\left( x_B \right) . 
\end{equation}

\noindent
The boson-fermion symmetry suggests that we should choose each term in (5.18)
to have the same form as (5.3).  We can also write

\begin{equation}
\Psi_r^B = U_B {\psi}_r^B = n_B^{1/2} U_B \stackrel{\sim}{\psi}_r^B 
\end{equation}

\noindent
as in (5.15), and choose the $\Psi_r^B$ to be eigenfunctions of the operator in
(5.14):

\begin{equation}
\left( -{1 \over {2m}} \partial^m \partial_m + V - \mu_B \right) \Psi_r^B =
\varepsilon_r \Psi_r^B . 
\end{equation}

\noindent
We will find, as usual, that only the solutions with $\varepsilon_r$ = 0 can be
retained in the low-energy limit.$^{3,33-35}$  The above choices and ideas are
similar to those of other higher-dimensional theories, and it will be seen that
they lead to consistent results.

The internal space $x_B$ has an effective geometry determined by the effective
metric tensor
$g_{mn}$.  One can then define Killing vectors $K_i^n$, or

\begin{equation}
K_i = K_i^n \partial_n . 
\end{equation}

\noindent
They have an algebra$^{14,36}$

\begin{equation}
K_i K_j - K_j K_i = - c_{ij}^k K_k 
\end{equation}

\noindent
and satisfy Killing's equation$^{15,37}$

\begin{equation}
K_i^p \partial_p g_{mn} + g_{pn} \partial_m K_i^p + g_{mp} \partial_n K_i^p = 0 .
\end{equation}

\noindent
For a scalar function $F$ which is invariant under the symmetry operation
specified by $K_i$, the corresponding equation is

\begin{equation}
K_i F = 0 . 
\end{equation}

We now need two assumptions: First, the condensate density $n_B$ is assumed to
have the same symmetry as the geometry defined by $g_{mn}$:

\begin{equation}
K_i n_B \left( x_B \right) = 0 .  
\end{equation}

\noindent
This will be the case if the velocity $v_m$ results from an instanton with
spherical symmetry which is ``frozen into'' internal space, as in the examples
of Section 7.  Second, the physically significant zero modes of (5.20) are
assumed to share this symmetry:

\begin{equation}
K_i \stackrel{\sim}{\psi}_r^B = 0 ,~\varepsilon_r = 0 . 
\end{equation}

\noindent
This assumption is reasonable because $\stackrel{\sim}{\psi}_r^B$, defined in
(5.19), satisfies the same equation as the constant vector $\eta_B$ of (5.15). 
It is also plausible that zero modes should reflect the symmetry of the space
in which they are defined.  A more detailed discussion of these modes is given
in Section 8.

In the simple picture mentioned below (5.2) and (5.25), the $K_i$ are
associated with rotations in {\it d} dimensions, and thus with the
symmetry group SO({\it d}).

Since the $\Psi_r^B$ serve as basis functions, let

\begin{equation}
<r|Q|s> = \int d^d x~\Psi_r^{B \dagger} Q \Psi_s^B 
\end{equation}

\noindent
where $Q$ is any operator.  In the next section we will need these functions to
be orthogonal,

\begin{equation}
<r|s> = \delta_{rs} 
\end{equation}

\noindent
and we will also need the result

\begin{equation}
K_i \Psi_s^B = U_B \left( imK_i^n v_n^c \sigma_c
\right) \psi_s^B \end{equation}

\noindent
which follows from (5.16), (5.25), and (5.26).  This implies the relation

\begin{equation}
\int d^d x~\psi_r^{B\dagger} \sigma_i \psi_s^B = <r|(-iK_i) |s> 
\end{equation}

\noindent
where

\begin{equation}
\sigma_i = mK_i^n v_n^c \sigma_c 
\end{equation}

\noindent
is a matrix associated with the $i^{th}$ internal symmetry direction.

\pagebreak
\section{Gauge fields}

In conventional Kaluza-Klein theories,$^{33,34}$ the metric tensor is perturbed
by letting

\begin{equation}
g_{\mu m} \left( x_A , x_B \right) = A_{\mu}^i \left( x_A \right) K_i^n \left(
x_B \right) g_{nm}. 
\end{equation}

\noindent
In the present theory, this corresponds to letting

\begin{equation}
v_{\mu}^c = A_{\mu}^i K_i^n v_n^c 
\end{equation}

\noindent
since $g_{\mu m} = v_{\mu}^c v_m^c$ (with $v_m^{\alpha}$ still zero) and
$g_{nm} = v_n^c v_m^c$.  It is also equivalent to writing

\begin{equation}
mv_{\mu}^c \sigma_c = A_{\mu}^i \sigma_i . 
\end{equation}

We now need to determine how the gauge fields $A_{\mu}^i$ are coupled to the
fermion fields $\psi_r$.  When (4.2) and (5.18) are substituted into (4.1), and
(5.20) is used (with $\varepsilon_r$ = 0), the result is

\begin{equation}
S_f = \sum_{rs} \int d^D x~h \psi_r^{B\dagger} \psi_r^{\dagger} U^{\dagger}
\left( - \frac{1}{2m} \partial^{\mu} \partial_{\mu} - \frac{1}{2}
mv_{\alpha}^{\mu} v_{\mu}^{\alpha} \right) U \psi_s \psi_s^B . 
\end{equation}

\noindent
Let us focus on the term involving $\partial^{\mu} \partial_{\mu}$, and a
particular $r$ and $s$, which becomes

\begin{equation}
S_{rs} = \frac{1}{2m} \int d^D x~h \psi_r^{B\dagger} \left[ \partial^{\mu} \left(
U \psi_r \right) \right]^{\dagger} \partial_{\mu} (U \psi_s) \psi_s^B 
\end{equation}

\noindent
after integration by parts.  Since (3.13) and (5.4) give

\begin{equation}
\partial_{\mu} U = imU \left( v_{\mu}^{\alpha} \sigma_{\alpha} + v_{\mu}^c
\sigma_c \right) 
\end{equation}

\noindent
we need to consider

\begin{equation}
\partial_{\mu} \left( U \psi_s \right) = U \left( imv_{\mu}^{\alpha}
\sigma_{\alpha} + \partial_{\mu} \right) \psi_s + Uimv_{\mu}^c \sigma_c \psi_s
. 
\end{equation}

\noindent
This expression is multiplied by its conjugate (with $s \rightarrow r$).  The
product of the first term with its conjugate was already treated in Section 4. 
The product of the second term with its conjugate is second order in
$v_{\mu}^c$, and can consequently be neglected.  The extra contribution to
(6.5) thus involves the cross terms $\left( im \psi_r \right) ^{\dagger}
v_c^{\mu} \sigma^c \left( imv_{\mu}^{\alpha} \sigma_{\alpha} + \partial_{\mu}
\right) \psi_s$ + (conj. with $r \leftrightarrow s$).  For low-energy
excitations, however, $\partial_{\mu} \psi_s$ can be neglected in comparison with
$mv_{\mu} \psi_s$.  We are left with

\begin{equation}
m^2 \psi_r^{\dagger} v_{\mu}^c \sigma_c v_{\alpha}^{\mu} \sigma^{\alpha} \psi_s
\nonumber
\end{equation}

\noindent
plus its conjugate.  The additional term in (6.5) is then

\begin{equation}
\Delta S_{rs} = \frac{1}{2} \int d^4 x~h \psi_r^{\dagger} v_{\alpha}^{\mu}
\sigma^{\alpha} \left[ \int d^d x~\psi_r^{B \dagger} mv_{\mu}^c \sigma_c
\psi_s^B \right] \psi_s + conj. 
\end{equation}

When (6.3) and (5.30) are employed, the factor in square brackets reduces to a
remarkably nice form:

\begin{eqnarray}
\Delta S_{rs}^{\mu} &=& \int d^d x~\psi_r^{B\dagger} mv_{\mu}^c \sigma_c
\psi_s^B \\
&=& A_{\mu}^i \int d^d x~\psi_r^{B\dagger} \sigma_i \psi_s^B \\
&=& A_{\mu}^i \langle r| \left( -iK_i \right) |s \rangle.
\end{eqnarray}

\noindent
Then (6.9) can be rewritten as

\begin{equation}
\Delta S_{rs} = \frac{1}{2} \int d^4 x~h \psi_r^{\dagger} v_{\alpha}^{\mu}
\sigma^{\alpha} A_{\mu}^i t_i^{rs} \psi_s + conj. 
\end{equation}

\noindent
where

\begin{equation}
t_i^{rs} = <r| \left( -iK_i \right) | s> . 
\end{equation}

\noindent
Let $t_i$ be the matrix with elements $t_i^{rs}$.  Since it corresponds to the
operator -i$K_i$, it has the same algebra:

\begin{equation}
t_i t_j - t_j t_i = ic_{ij}^k t_k . 
\end{equation}

\noindent
This is exactly what is needed for (6.13) to represent a proper gauge
interaction.

To simplify notation, let $\psi$ be the vector with components $\psi_r$.  Then
the extra contribution to (6.4) is

\begin{equation}
\Delta S = \frac{1}{2} \int d^4 x~h \psi^{\dagger} e_{\alpha}^{\mu}
\sigma^{\alpha} A_{\mu}^i t_i \psi +conj. 
\end{equation}

\noindent
After (5.28) is used, the other terms in (6.4) can be treated just as in
Section 4.  The Lagrangian density corresponding to (6.16) can then be added
to (4.11), giving

\begin{equation}
\bar{\cal L}_f = - \frac{1}{2} ih~\psi^{\dagger} e_{\alpha}^{\mu}
\sigma^{\alpha} \stackrel{\sim}{D}_{\mu} \psi + conj. 
\end{equation}

\noindent
where

\begin{equation}
\stackrel{\sim}{D}_{\mu} = \stackrel{\sim}{\nabla}_{\mu} + iA_{\mu}^i t_i .
\end{equation}

\noindent
The present theory thus yields the correct form for initially massless fermions
coupled to both gravity and gauge fields.

\pagebreak
\section{Instantons}

In an ordinary superfluid, the definition m$\vec{v}$ =
${\vec{\nabla}} \theta$ implies that

\begin{equation}
\vec{\nabla} \times \vec{v} = 0. 
\end{equation}

\noindent
For the condensate of the present theory, the definition $mv_M = - iU^{-1}
\partial_M U$, together with the condition $\partial_M \partial_NU$ - $\partial_N
\partial_M U = 0$, immediately gives the generalization

\begin{equation}
G_{MN} = 0 
\end{equation}

\noindent
where

\begin{equation}
G_{MN} = \partial_M v_N - \partial_N v_M + im \left[ v_M , v_N \right] .
\end{equation}

\noindent
If (7.2) were to hold everywhere, the present theory would be untenable, since
there is no such constraint on the vielbein and metric tensor in standard
physics.  It seems to be a general principle, however, that constraints like
(7.2) can be relieved by topological defects, with important physical
consequences.  Let us consider a few examples.

(a) \underline{U(1) vortices}.  Since (7.1) states that
$\vec{v}$ is irrotational, it was originally a mystery how
superfluid $^4$He could exhibit its observed rotation.  Feynman provided an
answer by postulating the existence of vortices,$^{5}$ which were later seen
experimentally.  Integration over an area A containing a vortex gives

\begin{eqnarray}
\int_A \vec{\nabla} \times \vec{v} \cdot d
\vec{S} &=& \int_C \vec{v} \cdot d \vec{\ell} \nonumber \\
&=& 2 \pi n/m 
\end{eqnarray}

\noindent
where $n$ is an integer.  The singularity at the center of a vortex thus relieves
the constraint (7.1), in the sense that the integrated value of $\vec{\nabla}
\times \vec{v}$ in (7.4) is nonzero.  This has the important effect of allowing
the superfluid to rotate.

(b) \underline{SU(2) instantons in four dimensions}.  The velocity field around
an $n$=1 vortex is given by$^{5}$

\begin{equation} 
m\vec{v} (\vec{r}) = r^{-1}\hat{\phi} 
\end{equation}

\noindent
where $\vec{r}~=~(r,\phi)$ in the xy plane.  For an $n$=1 BPST instanton the
corresponding result is$^{12,32}$

\begin{equation}
m\vec{v}(x)~=~\frac{\vec{\sigma} x_0 + \vec{\sigma} \times
\vec{x}}{\rho^2},~~~~~mv_0 (x)~=~\frac{\vec{\sigma} \cdot \vec{x}}{\rho^2} 
\end{equation} 

\noindent
where

\begin{equation}
\rho^2~=~x_0^2 + \vec{x}^2 . 
\end{equation}

\noindent
(The replacement $A_{\mu}$/i $\rightarrow mv_{\mu}$ has been made in the usual
expressions, and the instanton size $\lambda$ has been set equal to zero.) 
Even though (7.2) is satisfied at all points except $\rho$=0, the integrated
value of G$^{\mu \nu}$G$_{\mu \nu}$ is nonzero:

\begin{equation}
\int d^4 x~tr \left( G^{\mu \nu} G_{\mu \nu} \right) = 16 \pi^2 / m^2
. 
\end{equation}

(c)  \underline{Gravitational instantons in four dimensions}.  The
Eguchi-Hanson instanton has a metric which can be written in the form$^{28}$

\begin{eqnarray}
ds^2 = \left( 1 - a^4r^{-4} \right) ^{-1} dr^2 &+& \left( 1 - a^4r^{-4} \right)
\left( r^2 / 4 \right) \left( d \psi + cos\theta~d \phi\right) ^2 \nonumber \\
&+& \left( r^2 /4 \right) \left( d \theta^2 + sin^2 \theta~d \phi^2 \right) .
\end{eqnarray}

\noindent
As $r/a \rightarrow \infty$ this becomes the metric of flat Euclidean space,
but the Euler number

\begin{equation}
\chi = \left( 128 \pi^2 \right) ^{-1} \int d^4 x~g \varepsilon_{\mu
\nu}^{\tau \omega} R_{\tau \omega \alpha \beta} \varepsilon_{\rho
\sigma}^{\alpha \beta} R^{\mu \nu \rho \sigma}~+~boundary~terms 
\end{equation}

\noindent
and the signature

\begin{equation}
\tau~=~ \left( 96 \pi^2 \right) ^{-1} \int d^4 x~g R_{\mu \nu \rho \sigma}
\varepsilon^{\rho \sigma \tau \omega} R^{\mu \nu}~_{\tau \omega}~+~boundary~terms
\end{equation}

\noindent
are nonzero:  $\chi$=2 and $\tau$=1.  Gibbons and Hawking obtained a
generalization with the form$^{28}$

\begin{equation}
ds^2~=~u^{-1} \left( d\tau + \vec{\omega} \cdot d\vec{x} \right) ^2 + u~
d\vec{x} \cdot d\vec{x} 
\end{equation}

\noindent
where

\begin{equation}
u = u_0 + \sum_{i=1}^s q_i | \vec{x} - \vec{x}_i | ^{-1} 
\end{equation}

\begin{equation}
\vec{\nabla} \times \vec{\omega} = \vec{\nabla} u . 
\end{equation}

\noindent
When $u_0$ = 0 and all the $q_i$ are equal, these solutions are also
asymptotically locally Euclidean, but with $\chi$=s and $\tau$=s-1.

(d) \underline{Multidimensional instantons}.  The above ideas are known to
generalize to larger symmetry groups and higher dimensions.$^{38-40}$  For
example, the Kaluza-Klein monopole in 5 dimensions is given by$^{41}$

\begin{equation}
ds^2 = dt^2 + u^{-1} \left( d\tau + \vec{\omega} \cdot d\vec{x} \right) ^2 +
u~d\vec{x} \cdot d\vec{x}
\end{equation}

\noindent
where the fifth coordinate $\tau$ is periodic and $u_0$ = 1, s = 1 in (7.13).

(e) \underline{Other topological defects in field theory}$^{12,28,42-48}$ which
play an important role in grand-unified theories, higher-dimensional theories,
and cosmological models.

(f) \underline{Other topological defects in condensed-matter physics},$^{6-11}$
which have a pervasive influence on the properties of superfluid $^3$He and
$^4$He, type I and type II superconductors, liquid crystals, crystalline
solids, magnetic materials, one-dimensional organic systems, and
two-dimensional phase transitions.

Given the ubiquity of topological defects, it is not unreasonable to assume
that they exist in an ordered phase of the kind proposed here.  The vortices
postulated by Feynman relaxed the constraint (7.1), permitting the integrated
vorticity to be nonzero.  The topological defects postulated here will
similarly relax the constraint (7.2), permitting integrated curvature scalars
like (7.8) to be nonzero.

Three distinct kinds of defects are needed:

(1) \underline{An internal instanton}, associated with the symmetry group G,
which accounts for the internal velocity field $v_m^c$.  This topological
defect is analogous to the monopoles, instantons, etc. which are postulated in
other higher-dimensional theories.  Since G is left unspecified in the present
paper, so is the detailed nature of this instanton.  The simplest toy models
are the following:

\indent
(a) \underline{d $\rightarrow$ 2 and G $\rightarrow$ U(1)}.  Then the
condensate is bounded by a circle of radius $r_B$ in internal space, and
$V_B~=~\pi r_B^2$.  The internal instanton is a vortex, with $v_\phi =
(mr)^{-1}$.  There is only one Killing vector

\begin{equation}
K~=~\partial_{\phi} 
\end{equation}

\noindent
and the gauge group is U(1).

\indent
(b) \underline{d $\rightarrow$ 4 and G $\rightarrow$ SU(2)}.  In this case the
condensate is enclosed by a 3-sphere of radius $r_B$.  The internal instanton
has the form (7.6).  There are now 6 Killing vectors $K_i$, associated with the
6 rotational degrees of freedom, and the gauge group is SO(4).

A more realistic model is provided by \underline{d $\rightarrow$ 10}.  Then the
condensate lies within a 9-sphere of radius $r_B$, and the internal instanton
is a hypothetical extension of (7.6) to a larger symmetry group which is
contained in G.  There are 45 Killing vectors and the gauge group is SO(10),
perhaps the most appealing possibility for grand unification.$^{49}$

A finite internal volume $V_B$ is required for this instanton to have finite
action: If $mv_m^c \propto r^{-1}$, as in (7.5) and (7.6), then the kinetic
energy contribution

\begin{equation}
\int d^d x~n_B \cdot \frac{1}{2} mv_c^m v_m^c 
\end{equation}

\noindent
will diverge unless $n_B \rightarrow$ 0 for $r > r_B$.  Since the natural
length scale in (5.14) is the correlation length$^{5}$

\begin{equation}
\xi = \left( 2 m\mu \right) ^{-1/2} 
\end{equation}

\noindent
it is plausible that

\begin{equation}
r_B \sim \xi ,~~~V_B \sim \xi^d . 
\end{equation}

\noindent
Notice that the internal velocity $v_m$ has no radial component.  The metric
tensor $g_{mn}$ is then defined only along the tangential directions, with
$v_m^c \propto \left( mr \right) ^{-1}$ and $g_{mn} \propto \left( mr \right)
^{-2}$ within a ($d$-1)-sphere of radius $r$.  Let $V_B^{\prime}$ be the
effective volume of this sphere:

\begin{equation}
V_B^{\prime} = \int d^{d^{\prime}} x~g_{d^{\prime}}~~~, d^{\prime} = d - 1
\end{equation}

\noindent
where

\begin{equation}
g_{d^{\prime}} = \left( det~g_{mn} \right) ^{1/2} 
\end{equation}

\noindent
and

\begin{center}
$m,n$ = 4,5, $\ldots$, 3 + $d^{\prime}$.
\end{center}

\noindent
We can similarly define

\begin{equation}
g_{D^{\prime}} = \left( det~g_{MN} \right) ^{1/2} , 
\end{equation}

\noindent
where $D^{\prime} = D$-1 and the coordinates are restricted to those describing
the manifold $R^4 \times S^{d^{\prime}}$:

\begin{center}
$M$,$N$ = 0,1, $\ldots$, $D^{\prime}$ -1 .
\end{center}

\noindent
Since $g_{d^{\prime}} \propto \left( mr \right) ^{-d^{\prime}}$ and
d$^{d^{\prime}} x = r^{d^{\prime}}$ d$\Omega$, where d$\Omega$ is a solid
angle, (7.20) implies that $V_B^{\prime}$ is independent of $r$ and

\begin{equation}
V_B^{\prime} \sim m^{-d^{\prime}} . 
\end{equation} 

(2) \underline{A cosmological instanton}, with an SU(2) velocity field like
that of (7.6).  If we choose $\vec{x}$ = 0 at our position in the universe,
then the 3-vector $v_k^a \sigma_a$ has the form

\begin{equation}
\vec{v} \propto \vec{\sigma} / m x^0 
\end{equation}

\noindent
and $v_0^a$ = 0.  The singularity at $x^0$ = 0 is interpreted as the big bang. 
Recall that there is also a U(1) velocity field $v_{\mu}^0$, which need not be
real, and that $\Psi$ and $\Psi^{\dagger}$ vary independently.  These features
can be exploited in minimizing the action (3.3), by requiring the U(1) kinetic
energy $\frac{1}{2} mv_0^{\mu} v_{\mu}^0$ to be negative.  Symmetry indicates
that $v_{\mu}^0$ is radial, or along the $x^0$ direction at our position in the
universe, giving the texture (4.7).  $\Psi_s$ then varies as exp$\left( -\omega
x^0 \right)$ within the present Euclidean picture, with $\Psi_s^{\dagger} \propto
exp \left( +\omega x^0 \right)$ to keep $n_s$ constant.  In Section 9 we will
transform to a Lorentzian picture by performing a Wick rotation $x^0
\rightarrow ix^0$.  The above dependences are then changed to exp$\left(
-i\omega x^0 \right)$ and exp$\left( +i\omega x^0 \right)$, with the condensate
density still constant.

The continuity equation (3.16) appears to impose a constraint on the velocity
field $v_{\mu}^0$, but this constraint may also be relieved by topological
defects: There can be monopole-like defects which act as sources or sinks for
the current $j_{\mu}^0 = \eta^{\dagger} n_s v_{\mu}^0 \eta$, with
$\partial^{\mu} j_{\mu}^0$ = 0 everywhere except at the singularities themselves
(where $n_s \rightarrow$ 0).  These defects are physically allowed because
$v_{\mu}^{\alpha}$ is not a true superfluid velocity; it instead specifies a
field configuration, analogous to the configuration of spins on a lattice.

(3) \underline{Planck-scale instantons} which are dilutely distributed
throughout external spacetime, and which give rise to a twisting of the field
$e_{\mu}^A$.  Just as an ordinary gravitational instanton
is embedded in a surrounding metric $g_{\mu \nu}$, or vierbein
$e_{\mu}^{\alpha}$, the instantons postulated here are embedded in a more
general field $e_{\mu}^A$ which includes both the gravitational field (for $A$ =
$\alpha \leq$ 3) and the gauge fields (for $A$ = c $\geq$ 4).

The effective vielbein $e_M^A$ and metric tensor $g_{MN}$ of (5.8) are defined
on a manifold ${\cal M}$ of dimension $D^{\prime}$.  (Recall that
$D^{\prime}$ = $D$ - 1 and ${\cal M} = R^4 \times S^{D^{\prime} - 4}$ in the models
above (7.23).)  We can then define a Riemannian curvature scalar
$^{(D^{\prime})}R$ and a scalar density $g_{D^{\prime}}$ = $det~g_{MN}$, with
the coordinates $M$ and $N$ restricted to this manifold.

Let $S_{in}$ be the action of one instanton, and $R_{in}$ be its contribution to
the quantity

\begin{equation}
-\int d^{D^{\prime}}x~g_{D^{\prime}}~^{(D^{\prime})} R . 
\end{equation}

\noindent
(It is assumed that each instanton has a core singularity which makes $R_{in}$
nonzero.  It is also assumed that instantons of the same kind have the same
values of $S_{in}$ and $R_{in}$.)  A comparison of (7.29) with (7.30) shows that
$R_{in}$ must be positive.

Although it costs an action $S_{in}$ to form an instanton, the action of the
matter fields can be lowered by the resulting change in curvature.  We will
find below that minimization of the total action with respect to $g^{\mu
\nu}$ and $A_{\mu}^i$ leads to the Einstein and Maxwell field equations.

Since $S_{in}$ is dimensionless and $^{(D^{\prime})} R$ has dimension
length$^{-2}$, we can write

\begin{equation}
S_{in} = \ell_0^{-D^{\prime} + 2} R_{in} . \end{equation}

\noindent
To obtain precise values of $S_{in}$, $R_{in}$, and $\ell_0$ would require
detailed calculations for a specific model.  We can, however, obtain estimates
if the instantons are assumed to have the following general properties:  First,
the presence of $m$ in (3.13) suggests a ``velocity core'' of size $r_v \sim
m^{-1}$, within which $v_{\mu}^A \sim$ 1 and $\partial_{\mu} v_{\nu}^A \sim m$. 
(This behavior can be seen explicitly in (7.5) and (7.6), which become
dimensionless if distances are scaled by $m^{-1}$.)  Second, the presence of a
singularity suggests a ``density core'' of size $r_n \sim \xi$.  One then
expects $R_{in} \sim m^{-4} V_B^{\prime} m^2$ and $S_{in} \sim \xi^4 V_B \mu^2 /
b$.  (The action (3.3) becomes $\int d^D x \left( -\frac{1}{2} bn_s^2 \right)$
after (3.7) is used.  If $\Psi_s$ is constant, (3.7) also implies that the
density is $\bar{n}_s = \mu/b$.  Then in a core region of radius $r_n$, whose
density is depleted by a singularity at the center, the change in the action is
$\Delta S = \int d^D x \left( - \frac{1}{2} bn_s^2 + \frac{1}{2} b \bar{n}_s^2
\right) \,\sim\,\xi^4 V_B \mu^2$ /b.)  It follows that

\begin{equation}
\ell_0^{D^{\prime} - 2} \sim \xi^{-d} V_B^{\prime} b \sim (\mu / m) ^{d/2} mb
\end{equation}

\noindent
where (7.18), (7.19), and (7.23) have been used.  Finally, (7.31) relates the
Planck length $\ell_P$ to the parameters m, $\mu$, and b of (2.7):

\begin{equation}
\ell_P^2 \sim (m\mu)^{d/2} b,~~~d = D - 4. 
\end{equation}

Since the contributions are additive for dilutely distributed instantons,
(7.26) implies that they have a net action

\begin{equation}
S_{D^{\prime}} = - \ell_0^{-D^{\prime} + 2} \int d^{D^{\prime}}x~
g_{D^{\prime}}~^{(D^{\prime})} R_{in} 
\end{equation}

\noindent
where $^{(D^{\prime})}R_{in}$ represents their total contribution to the
scalar curvature.  This is the Euclidean Einstein-Hilbert action in D$^{\prime}$
dimensions, and the usual Kaluza-Klein reduction gives$^{33,34}$

\begin{equation}
S_{D^{\prime}} = -\ell_P^{-2} \int d^4 x~g~^{(4)} R + {1 \over 4} g_0^{-2} \int
d^4 x~g F_{\mu \nu}^i F_{\rho \sigma}^i g^{\mu \rho} g^{\nu \sigma} 
\end{equation}

\noindent
where

\begin{equation}
\ell_P^{-2} = \ell_0^{-D^{\prime} + 2} V_B^{\prime} 
\end{equation}

\begin{equation}
\ell_P^{-2} < g_{mn} K_i^m K_j^n > = g_0^{-2} \delta_{ij} 
\end{equation}

\noindent
and

\begin{equation}
F_{\mu \nu}^i = \partial_{\mu} A_{\nu}^i - \partial_{\nu} A_{\mu}^i
+ c_{jk}^i A_{\mu}^j A_{\nu}^k . 
\end{equation}

\noindent
$V_B^{\prime}$ is the internal volume of (7.20) and $<$ --- $>$ represents an
average over this volume.  (If $\sigma_i$ is constant in (5.31), however, then
so is $g_{mn} K_i^m K_j^n$, eliminating the need for an average.)  As in
conventional Kaluza-Klein theories, $^{(4)}R$ is the curvature scalar associated
with the vierbein $e_{\mu}^{\alpha}$, and $g = \left( det~e_{\mu}^{\alpha}
e_{\nu}^{\alpha} \right) ^{1/2}$.

Suppose that $v_m^c v_n^c K_i^m K_j^n$ is $\sim m^{-2} \delta_{ij}$, as in the
models above (7.23).  Since $g_0$ is $\sim$ 1,$^{12,13}$ (7.32) then implies the
relationship

\begin{equation}
m \sim m_P . 
\end{equation}

\noindent
The Lagrangian densities corresponding to (7.30) are

\begin{equation}
\bar{\cal L}_G = -\ell_P^{-2} g~^{(4)} R \end{equation}

\begin{equation}
\bar{\cal L}_g = \frac{1}{4} g_0^{-2} g F_{\mu \nu}^i F_{\rho
\sigma}^i g^{\mu \rho} g^{\nu \sigma} . 
\end{equation}

\pagebreak
\section{Bosonic excitations}

When bosonic excitations $\Phi_b$ are included, (2.11) becomes

\begin{equation}
S = S_b + S_f 
\end{equation}

\begin{equation}
S_b = \int d^Dx~h \Psi_b^{\dagger} \left( T + {1 \over 2} \stackrel{\sim}{V} - \mu
\right) \Psi_b 
\end{equation}

\begin{equation}
S_f = \int d^Dx~h \Psi_f^{\dagger} \left( T + \stackrel{\sim}{V} - \mu \right) \Psi_f
\end{equation}

\noindent
where

\begin{equation}
\stackrel{\sim}{V} = b \Psi_b^{\dagger} \Psi_b 
\end{equation}

\noindent
and $\Psi_b = \Psi_s + \Phi_b$.  If we now require that

\begin{equation}
\left( T + \stackrel{\sim}{V} - \mu \right) \Psi_s = 0 
\end{equation}

\noindent
the treatment of the order parameter and fermionic excitations in the
preceding sections is unchanged, except that $V \rightarrow
\stackrel{\sim}{V}$.  The bosonic action (8.2) can be written

\begin{equation}
S_b = S_0 + \Delta S_b + \Delta S_b^{\prime} 
\end{equation}

\noindent
with

\begin{equation}
\Delta S_b = \int d^D x~h \Phi_b^{\dagger} \left( T + V - \mu + \frac{1}{2} b
\Phi_b^{\dagger} \Phi_b \right) \Phi_b 
\end{equation}

\begin{equation}
\Delta S_b^\prime = \int d^D x~h \Phi_b^{\dagger} \left( T + \frac{1}{2}
\stackrel{\sim}{V} + \frac{1}{2} V + \frac{1}{2} b \Phi_b^{\dagger} \Phi_b - \mu \right)
\Psi_s + conj. 
\end{equation}

\noindent
For the excitations considered below, we will find that $\Phi_b^{\dagger}
\Psi_s$ = 0, so that (8.8) is unchanged if the interaction terms in parentheses
are replaced by $\stackrel{\sim}{V}$.  But the equation of motion (8.5) then gives

\begin{equation}
\Delta S_b^{\prime} = 0 . 
\end{equation}

Let us expand the boson field $\Phi_b$ in the complete set of internal states
$\Psi_r^B$, with coefficients $\Phi_r$:

\begin{equation}
\Phi_b = \sum_r~^{\prime}~~\Phi_r \Psi_r^B 
\end{equation}

\noindent
where

\begin{equation}
\Psi_0^B = N_B^{-1/2} \Psi_B , 
\end{equation}

\noindent
$N_B = \int d^d x~n_B$, and the prime means r $\neq$ 0.  Recall that these basis
functions are the solutions to (5.20), and are written in the form (5.19).  In
treating low-energy bosonic excitations, it is necessary to assume the
orthogonality condition

\begin{equation}
\stackrel{\sim}{\psi}_r^{B \dagger} \stackrel{\sim}{\psi}_s^B = N_B^{-1}
\delta_{rs}~~~, \varepsilon_r = \varepsilon_s = 0 
\end{equation}

\noindent
or equivalently

\begin{equation}
\Psi_r^{B \dagger} \Psi_s^B = N_B^{-1} n_B \delta_{rs}~~~, \varepsilon_r =
\varepsilon_s = 0 . 
\end{equation}

\noindent
Only those functions satisfying this condition and (5.26) are considered to be
physically significant in the present context.  There is another set of solutions
to (5.20) with $\Psi_r^B \rightarrow \left( \Psi_r^B \right) ^*$ ; since these
involve motion counter to that of the condensate, however, it is assumed that
radiative corrections will break the degeneracy between these states and those of
(8.13), so they are omitted from the sums of (5.18) and (8.10) at low energy. 
The state with r=0 is already occupied by the order parameter, so it is also
omitted from (8.10).  Then (5.9) and (8.10) -- (8.13) imply that

\begin{equation}
\Phi_b^{\dagger} \Psi_s = 0 
\end{equation}

\noindent
and

\begin{equation}
\Phi_b^{\dagger} \Phi_b = \sum_r~^{\prime}~~\Phi_r^{\dagger} \Phi_r \Psi_r^{B
\dagger} \Psi_r^B . 
\end{equation}

\noindent
Since $T = (2m)^{-1} \left( \partial^{\mu} \partial_{\mu} + \partial^m
\partial_m \right)$, the first term in (8.7) involves

\begin{eqnarray}
\partial_{\mu} \Phi_b &=& \partial_{\mu} \sum_s~^{\prime}~~\Phi_s U_B
\psi_s^B \\
&=& \sum_s~^{\prime}~~U_B \left( \psi_s^B \partial_{\mu} \Phi_s +
imv_{\mu}^c \sigma_c \psi_s^B \Phi_s \right)  \\
&=& \sum_s~^{\prime}~~\left( \Psi_s^B \partial_{\mu} \Phi_s + A_{\mu}^i K_i
\Psi_s^B \Phi_s \right) 
\end{eqnarray}

\noindent
where (8.10), (5.19), (6.6), (6.3), (5.31), and (5.29) have been used.  After
integration by parts, the $\partial^{\mu} \partial_{\mu}$ term of (8.7) then has
the form

\begin{equation}
\Delta S_1 = \int d^4 x~h(2m)^{-1} \sum_{rs}~^{\prime}~~\int d^d x \left(
P_r^{\mu} \Psi_r^B \right) ^{\dagger} \left( P_{\mu s} \Psi_s^B 
\right) 
\end{equation}

\noindent
where

\begin{equation}
P_{\mu s} = \partial_{\mu} \Phi_s + \Phi_s A_{\mu}^i K_i . 
\end{equation}

\noindent
The integral over the internal coordinates can be written

\begin{eqnarray}
\langle r | P_r^{\mu \dagger} P_{\mu s} | s \rangle &=& \sum_t \langle r | P_r^{\mu \dagger} | t \rangle \langle
t | P_{\mu s} | s \rangle  \\
&=& \sum_t \left( \delta_{tr} \partial^{\mu} \Phi_r + i A_i^{\mu} t_i^{tr} \Phi_r
\right) ^{\dagger}  \left( \delta_{ts} \partial_{\mu} \Phi_s + i A_{\mu}^j
t_j^{ts} \Phi_s \right) 
\end{eqnarray}

\noindent
after (6.14) is used.  Then (8.19) becomes

\begin{equation}
\Delta S_1 = \int d^4 x~h (2m)^{-1} D^{\mu} \Phi^{\dagger} D_{\mu} \Phi
\end{equation}

\noindent
where $\Phi$ is the vector with components $\Phi_r$ and

\begin{equation}
D_{\mu} = \partial_{\mu} + i A_{\mu}^i t_i . 
\end{equation}

\noindent
Notice that the bosons of this section have not been treated in the same way as
the fermions of Sections 4--6.  This is because the bosons can undergo
condensation at low energy.  Their equation of motion is then less important than
their coupling to the gauge fields $A_{\mu}^i$, and it is appropriate to deal
directly with the boson field $\Phi_b$ rather than writing it in the form (4.2)
and neglecting terms that are second order in $A_{\mu}^i$.

With $\varepsilon_r$ = 0, (5.20) and (8.10) imply that 

\begin{equation}
\left( - {1 \over {2m}} \partial^m \partial_m + V - \mu_B \right) \Phi_b = 0
\end{equation}

\noindent
so the next term from (8.7) is 

\begin{equation}
\Delta S_2 = - \int d^4 x~h \Phi^{\dagger} \mu_A \Phi . 
\end{equation}

\noindent
Also, (8.15) gives

\begin{equation}
\int d^d x \left( \Phi_b^{\dagger} \Phi_b \right) ^2 = \sum_{rs}~^{\prime}~~
\Phi_r^{\dagger} \Phi_r I_{rs} \Phi_s^{\dagger} \Phi_s 
\end{equation}

\noindent
where

\begin{equation}
I_{rs} = \int d^d x~\Psi_r^{B \dagger} \Psi_r^B \Psi_s^{B \dagger} \Psi_s^B .
\end{equation}

\noindent
For the solutions of (8.13), however, this expression is independent of $r$ and
$s$:  $I_{rs} = I$.  The last term from (8.7) is then

\begin{equation}
\Delta S_3 = \frac{1}{2} b I \int d^4 x~h \left( \Phi^{\dagger} \Phi \right) ^2
. 
\end{equation}

To obtain a standard form, let

\begin{equation}
\phi = (2m) ^{-1/2} \Phi 
\end{equation}

\begin{equation}
\bar{\mu}^2 = 2m \mu_A . 
\end{equation}

\noindent
The total Lagrangian density resulting from (8.23), (8.26), and (8.29) becomes

\begin{equation}
\bar{\cal L}_b = h \left[ D^{\mu} \phi^{\dagger} D_{\mu} \phi - \bar{\mu}^2
\phi^{\dagger} \phi + \frac{1}{2} \bar{b} \left( \phi^{\dagger} \phi \right) ^2
\right] 
\end{equation}

\noindent
where

\begin{equation}
\bar{b} = (2m)^2 b I . 
\end{equation}

The prefactor in (8.32) is $h = \left( det~h_{\mu \nu} \right) ^{1/2}$
rather than $g = \left( det~g_{\mu \nu} \right) ^{1/2}$, and the first term
involves

\begin{equation}
D^{\mu} \phi^{\dagger} D_{\mu} \phi = h^{\mu \nu} D_{\mu} \phi^{\dagger}
D_{\nu} \phi 
\end{equation}

\noindent
rather than $g^{\mu \nu} D_{\mu} \phi^{\dagger} D_{\nu} \phi$. 
Suppose for simplicity that $v_a^k = \lambda \delta_a^k$ and $v_0^0 = i
\lambda$ (with $v_a^0 = v_0^k$ = 0), since a similar scaling is implied by the
cosmological model of Section 7.  It follows that 

\begin{equation}
g^{\mu \nu} = \lambda^2 \delta^{\mu \nu} , g = \lambda^{-4} . 
\end{equation}

\noindent
Letting

\begin{equation}
\phi^{\prime} = \lambda \phi 
\end{equation}

\noindent
we can write

\begin{equation}
\bar{\cal L}_b = g \left[ g^{\mu \nu} D_{\mu} \phi^{\prime \dagger} D_{\nu}
\phi^{\prime} - \lambda^2 \bar{\mu}^2 \phi^{\prime \dagger} \phi^{\prime} +
{1 \over 2} \bar{b} \left( \phi^{\prime \dagger} \phi^{\prime} \right) ^2
\right] . 
\end{equation}

\noindent
We can similarly rescale (6.17):

\begin{equation}
\bar{\cal L}_f = - {1 \over 2} ig \psi^{\prime \dagger} e_{\alpha}^{\mu}
\sigma^{\alpha} \stackrel{\sim}{D}_{\mu} \psi^{\prime} + conj. 
\end{equation}

\noindent
where

\begin{equation}
\psi^{\prime} = \lambda^2 \psi . 
\end{equation}

The specific scaling of the preceding paragraph is simplistic, but it suggests
that the second term in (8.37) may be neglected, leaving

\begin{equation}
\bar{\cal L}_b = h \left[ D^{\mu} \phi^{\dagger} D_{\mu} \phi + {1 \over 2}
\bar{b} \left( \phi^{\dagger} \phi \right) ^2 \right] . 
\end{equation}

\noindent
(There is another reason for neglecting this term:  If $\mu_B$ is constant in
(5.14), $\mu_A$ must also be constant, and it is asymptotically equal to zero
in the cosmological picture of Section 7.  It follows that $\bar{\mu}$ = 0.) 
The final Lagrangian for fundamental bosons then has no mass terms or Yukawa
couplings.  This is consistent with the idea that radiative effects may give
rise to such additional interactions at the electroweak scale. On the other
hand, symmetry-breaking at a grand-unified scale is attributed to formation of
the order parameter itself: The argument that led to (8.23) and (8.29)
also implies that

\begin{equation}
S_0 = \int d^4 x~h \left[ (2m) ^{-1} D^{\mu} \stackrel{\sim}{\Phi}^{\dagger} D_{\mu}
\stackrel{\sim}{\Phi} - \mu N_B n_A + \frac{1}{2} (2m)^{-2} \bar{b} \left( N_B n_A
\right) ^2 \right] 
\end{equation}

\noindent
where $\stackrel{\sim}{\Phi}$ is the vector corresponding to $\Psi_s$, with all its
components $\stackrel{\sim}{\Phi}_r$ equal to zero except $\stackrel{\sim}{\Phi}_0 = N_B^{1/2}
\Psi_A$.  The gauge fields $A_{\mu}^i$ that are coupled to $\Psi_s$, through
the term $\stackrel{\sim}{\Phi}^{\dagger} \left( A_i^{\mu} t_i A_{\mu}^j t_j
\right) \stackrel{\sim}{\Phi}$ in (8.41), will acquire large masses when
$\Psi_s$ becomes nonzero.  According to (8.19) and (5.30), these are the fields
for which

\begin{equation}
\int d^d x \left( K_i \Psi_0^B \right) ^{\dagger} \left( K_j \Psi_0^B \right) =
N_B^{-1} \int d^d x~n_B \eta_B^{\dagger} \sigma_i \sigma_j \eta_B 
\end{equation}

\noindent
is nonzero, where $\sigma_i$ is defined in (5.31).  For example, if the
$\sigma_i$ were proportional to the SU(3) Gell-Mann matrices $\lambda_i$,
$^{12}$ and if $\eta_B^{\dagger}$ were (0,0,1), then the gauge fields
corresponding to i = 4,5,6,7,8 would acquire masses at the grand-unified scale,
and those corresponding to i = 1,2,3 would not, leaving an unbroken SU(2) gauge
group at lower energy.  The true internal symmetry group G should, of course,
leave an unbroken gauge group SU(3) $\times$ SU(2) $\times$ U(1).

Notice that the Bernoulli equation (3.26) is unchanged when $mv_{\mu}^c
\sigma_c = A_{\mu}^i \sigma_i$ is introduced at low energy.  For those
$A_{\mu}^i$ which do not couple to $\eta_B$, $\eta^{\dagger} \sigma_i \sigma_j
\eta$ vanishes in (3.21).  But those which do couple have large masses, so they
do not appear at low energy.

The scaling above (8.35) is also relevant to the extra fields of (4.11) --
(4.14):  If $v_{\alpha}^{\mu} \sim \lambda$, (4.16) shows that $e_{\mu}^{\alpha}
\sim \lambda^{-1}$, or $v_{\mu}^{\alpha} \sim \lambda^2 e_{\mu}^{\alpha}$,
giving

\begin{equation}
a_k \sim \lambda^2 me_k^0 ,~~~b_0 \sim \lambda^2 me_0^a \sigma_a . 
\end{equation}

\noindent
There is then an extra coupling to gravity for spin-polarized fermions which
involves a mass $\lambda^2m$.

\pagebreak
\section{Observable consequences}

The low-energy Lagrangian

\begin{equation}
\bar{\cal L} = \bar{\cal L}_f + \bar{\cal L}_b + \bar{\cal L}_g
+ \bar{\cal L}_G 
\end{equation}

\noindent
still corresponds to Euclidean spacetime.  We now need to perform a Wick
rotation$^{4,12,24-28}$

\begin{equation}
x^0 \rightarrow ix^0 
\end{equation}

\noindent
to obtain the Lorentzian action

\begin{equation}
S_L = iS = \int d^4 x~{\cal L} 
\end{equation}

\noindent
where

\begin{eqnarray}
{\cal L} &=& {\cal L}_f + {\cal L}_b + {\cal L}_g + {\cal L}_G  \\
{\cal L}_f &=& {1 \over 2} if \psi^{\dagger} e_{\alpha}^{\mu} \sigma^{\alpha}
\stackrel{\sim}{D}_{\mu} \psi + h.c.  \\ 
{\cal L}_b &=& - f \left[ D^{\mu}
\phi^{\dagger} D_{\mu} \phi + \frac{1}{2} \bar{b} \left( \phi^{\dagger} \phi
\right) ^2 \right]  \\
{\cal L}_g &=& - \frac{1}{4} g_0^{-2} e F_{\mu \nu}^i F_{\rho \sigma}^i g^{\mu
\rho} g^{\nu \sigma}  \\
{\cal L}_G &=& \ell_P^{-2} e~^{(4)} R 
\end{eqnarray}

\begin{equation}
e = \left| det~e_{\mu}^{\alpha} \right| = \left( - det~g_{\mu \nu} \right)
^{1/2} 
\end{equation}

\begin{equation}
f = \left( - det~h_{\mu \nu} \right) ^{1/2} 
\end{equation}

\noindent
and

\begin{equation}
D_{\mu} = \partial_{\mu} + iA_{\mu}^i t_i . 
\end{equation}

\noindent
A$_0^i, e_0^{\alpha}$, etc. are now real-valued Lorentzian fields (see, e.g.,
p. 329 of Ref. 28), and the metric tensors $h_{\mu \nu}$ and $g_{\mu
\nu}$ have Lorentzian signature (--+++).

${\cal L}$ contains four terms, corresponding respectively to spin 1/2
fermions, scalar bosons, gauge fields, and the gravitational field.  It has the
same form as the Lagrangian postulated in standard fundamental physics, except
for several differences that it is now appropriate to discuss.

For the sake of generality, suppose that radiative effects give rise to
additional interaction terms and an effective Lagrangian

\begin{eqnarray}
{\cal L}_{eff} &=& {\cal L} - fu \left( \phi \right) + {\cal L}_{int} \\
{\cal L}_{int} &=& - {1 \over 2} f \gamma \psi^{\dagger} \phi \psi
+ h.c.  \\
&=& - {1 \over 2} f \sum_{rps} \gamma_{rps} \psi_r^{\dagger}
\phi_p \psi_s + h.c. 
\end{eqnarray}

\noindent
where $u(\phi)$ contains terms of the form $\pm \mu_p^2 \phi_p^{\dagger}
\phi_p$.  The complete matter field Lagrangian is then

\begin{equation}
{\cal L}_m = {\cal L}_f + {\cal L}_B + {\cal L}_{int} 
\end{equation}

\noindent
with ${\cal L}_B = {\cal L}_b - fu \left( \phi \right)$.  Since the
fermions and fundamental bosons described by ${\cal L}_m$ are defined on an
initially flat spacetime with metric tensor $h_{\mu \nu}$, this Lagrangian
does not contain a conventional factor $e = \left( - det~g_{\mu \nu}
\right) ^{1/2}$.  Instead it contains the nondynamical factor $f = \left( - det~
h_{\mu \nu} \right) ^{1/2}$:

\begin{equation}
{\cal L}_m = f \stackrel{\sim}{\cal L}_m . 
\end{equation}

The variational principle (3.2) also holds for the Lorentzian action $S_L$:

\begin{equation}
\delta S_L = 0 . 
\end{equation}

\noindent
In addition, it holds for variations in $g^{\mu \nu}$, or
$e_{\alpha}^{\mu}$ = $v_{\alpha}^{\mu}$, since these are equivalent to
variations in $\Psi_s$ or $\Psi_b$.  The Einstein field equations are given as
usual by $\delta S_L / \delta g^{\mu \nu}$ = 0.  With the present action

\begin{equation}
S_m = \int d^4 x~f \stackrel{\sim}{\cal L}_m 
\end{equation}

\noindent
they are

\begin{equation}
R_{\mu \nu} - {1 \over 2} g_{\mu \nu}~^{(4)}R = - \ell_P^2 e^{-1} f
\frac{\delta \stackrel{\sim}{\cal L}_m}{\delta g^{\mu \nu}} - \ell_P^2 e^{-1}
\frac{\delta {\cal L}_g}{\delta g^{\mu \nu}} 
\end{equation}

\noindent
since $\delta e / \delta g^{\mu \nu} = - {1 \over 2} g_{\mu \nu} e$ and
$\delta~^{(4)} R / \delta g^{\mu \nu}$ is effectively $R_{\mu \nu}$.$^{14,15}$ 
With the conventional matter field action

\begin{equation}
S_m^{\prime} = \int d^4 x~e \stackrel{\sim}{\cal L}_m^{\prime} 
\end{equation}

\noindent
they are instead

\begin{equation}
R_{\mu \nu} - {1 \over 2} g_{\mu \nu}~^{(4)}R = - \ell_P^2 \left(
\frac{\delta \stackrel{\sim}{\cal L}_m^{\prime}}{\delta g^{\mu \nu}} -
{1 \over 2} g_{\mu \nu} \stackrel{\sim}{\cal L}_m^{\prime} \right) - \ell_P^2
e^{-1} \frac{\delta {\cal L}_g}{\delta g^{\mu \nu}} . 
\end{equation}

\noindent
Let us now consider the consequences of this modification.

(i) \underline{The cosmological constant}.  In conventional physics, the vacuum
has a Lagrangian density e $\stackrel{\sim}{\cal L}_0$ due to Higgs
fields.$^{50}$  If $\delta \stackrel{\sim}{\cal L}_0 / \delta g^{\mu \nu}$
is neglected, this density gives a contribution ${1 \over 2} \ell_P^2 g_{\mu
\nu} \stackrel{\sim}{\cal L}_0$ in the field equations (9.21), so it corresponds
to an effective cosmological constant $\Lambda$:$^{14,15}$

\begin{equation}
\Lambda = -{1 \over 2} \ell_P^2 \stackrel{\sim}{\cal L}_0 . 
\end{equation}

\noindent
This prediction of conventional physics is in error by at least 50 orders of
magnitude.$^{50}$  In the present theory, however, the Lagrangian density is $f
\stackrel{\sim}{\cal L}_0$, and there is no contribution involving
$\stackrel{\sim}{\cal L}_0$ directly in the field equations (9.19):

\begin{equation}
\Lambda = 0 . 
\end{equation}

\noindent
There may be a much weaker term involving $\delta \stackrel{\sim}{\cal L}_0 /
\delta g^{\mu \nu}$, but this appears to be consistent with observation. 
There is also a more poorly defined contribution due to vacuum 
fluctuations which is not considered here.

(ii) \underline{Ordinary matter as a gravitational source}.  Since
${\cal L}_B$ does not contribute in the field equations (9.19), we are left
with

\begin{eqnarray}
{\cal L}_F &=& f \stackrel{\sim}{\cal L}_F = {\cal L}_f + {\cal L}_{int} \nonumber \\
&=& {1 \over 2} f \psi^{\dagger} \left( ie_{\alpha}^{\mu} \sigma^{\alpha}
\stackrel{\sim}{D}_{\mu} - \gamma \phi \right) \psi + h.c. 
\end{eqnarray}

\noindent
The variational principle (9.17), for arbitrary $\delta \psi^{\dagger}$, then
gives the Dirac equation for initially massless fermions coupled to gauge fields
and scalar bosons:

\begin{equation}
\left( ie_{\alpha}^{\mu} \sigma^{\alpha} \stackrel{\sim}{D}_{\mu} - \gamma \phi \right)
\psi = 0. 
\end{equation}

\noindent
But this makes ${\cal L}_F$ = 0.  The same reasoning applies to the
corresponding action ${\cal L}_F^{\prime}$ in conventional physics, so the
conventional field equations (9.21) reduce to

\begin{equation}
R_{\mu \nu} - {1 \over 2} g_{\mu \nu}~^{(4)}R = -\ell_P^2
\frac{\delta \stackrel{\sim}{\cal L}_F^{\prime}}{\delta g^{\mu \nu}} - \ell_P^2
e^{-1} \frac{\delta {\cal L}_g}{\delta g^{\mu \nu}} . 
\end{equation}

\noindent
In ${\cal L}_F^{\prime}$, the fermion field $\psi^{\prime}$ has a normalization

\begin{equation}
\int d^4 x~e \psi^{\prime \dagger} \psi^{\prime} = N_f 
\end{equation}

\noindent
where $N_f$ is the total number of fermions.  In ${\cal L}_F$, on the other
hand, $\psi$ has a normalization

\begin{equation}
\int d^4 x~f \psi^{\dagger} \psi = N_f . 
\end{equation}

\noindent
When this difference is taken into account, the conventional field equations
(9.26) and the present field equations (9.19) make nearly the same
predictions.  It appears that both are in agreement with the classic and more
recent tests of general relativity.$^{14,15,51,52}$

As mentioned below (8.43), however, there is an extra coupling of fermions to
gravity through the fields $a_k$ and $b_0$, which might be observable.

(iii) \underline{Massless vector bosons}.  For photons and gluons the only
coupling to gravity is through the Lagrangian ${\cal L}_g$ of (9.7), and
this is the same in the present theory as in coventional physics.

(iv) \underline{Massive vector bosons}.  For the W and Z particles there is an
additional term resulting from (9.6).  In the present theory it is
\medskip
\begin{center}
$-fh^{\mu \nu} \left( A_{\mu}^i t_i \phi \right) ^{\dagger} \left(
A_{\nu}^j t_j \phi \right)$ .
\end{center}
\medskip
\noindent
whereas in conventional physics it would have the form
\medskip
\begin{center}
$-eg^{\mu \nu} \left( A_{\mu}^i t_i \phi^{\prime} \right) ^{\dagger} \left(
A_{\nu}^j t_j \phi^{\prime} \right)$ .
\end{center}
\medskip
\noindent
There is thus a difference in the coupling of these particles to gravity.  The
resulting violation of the equivalence principle will be small, because virtual
W-bosons have large masses, but it is potentially observable.

In addition to the above gravitational effects, the present theory predicts
unconventional behavior of propagators at high energy:  For $p^{\mu} >
mv^{\mu}$, the approximation below (4.4) will fail, and fermion propagators
should begin to go as $p^{-2}$ rather than $\rlap/p^{-1}$.  Also, the equation of
motion for scalar bosons involves $h^{\mu \nu} \partial_{\mu}
\phi^{\dagger} \partial_{\nu} \phi$ rather than $g^{\mu \nu} \partial_{\mu} 
\phi^{\prime
\dagger} \partial_{\nu} \phi^{\prime}$.  Since the model scaling above
(8.35) is not quantitatively correct, there will be a violation of Lorentz
invariance which should lead to observable effects for Higgs bosons.

Finally, the present theory provides a new cosmological picture, with
implications for the Hubble constant$^{53}$ and other large-scale properties 
of the universe.

\pagebreak
\noindent
{\bf Acknowledgements}
\\

I have greatly benefitted from discussions with R.L. Arnowitt, M.J. Duff, C.N.
Pope, E. Sezgin, S.A. Fulling, C.-R. Hu, and V. Pokrovsky.  This work was
supported by the Robert A. Welch Foundation.

\pagebreak
\noindent
{\bf References}
\\

\begin{enumerate}
\item
D. Bailin and A. Love, {\it Supersymmetric Gauge Field Theory and String
Theory} (Institute of Physics, Bristol, 1994).  Citations of the original
papers are given in this and the other books and articles listed below.

\item
{\it Supergravities in Diverse Dimensions}, edited by A. Salam and E.
Sezgin (North-Holland, Amsterdam, 1989).

\item
M.B. Green, J.H. Schwarz, and E. Witten, {\it Superstring Theory}
(Cambridge University Press, Cambridge, 1987).

\item
M. Le Bellac, {\it Quantum and Statistical Field Theory} (Clarendon,
Oxford, 1991).

\item
A.L. Fetter and J.D. Walecka, {\it Quantum Theory of Many-Particle
Systems} (McGraw-Hill, San Francisco, 1971).

\item
D.R. Tilley and J. Tilley, {\it Superfluidity and Superconductivity},
Third Edition (Adam Hilger, Bristol, 1990).

\item
D. Vollhardt and P. W\"{o}lfle, {\it The Superfluid Phases of Helium
3} (Taylor and Francis, London, 1990).

\item
P.W. Anderson, {\it Basic Notions of Condensed Matter Physics} (Benjamin,
Menlo Park, 1984).

\item
P.G. de Gennes, {\it Superconductivity of Metals and Alloys} (Benjamin,
Menlo Park, 1966).

\item
P.M. Chaikin and T.C. Lubensky, {\it Principles of Condensed Matter
Physics} (Cambridge University Press, Cambridge, 1995).

\item
N.D. Mermin, {\it Rev. Mod. Phys.} {\bf 51}, 591 (1979).

\item
T.-P. Cheng and L.-F. Li, {\it Gauge Theory of Elementary Particle
Physics} (Clarendon, Oxford, 1984).

\item
E. Leader and E. Predazzi, {\it An Introduction to Gauge Theories and
Modern Particle Physics} (Cambridge University Press, Cambridge, 1996).

\item
C.W. Misner, K.S. Thorne, and J.A. Wheeler, {\it Gravitation} (W.H.
Freeman, San Francisco, 1973).

\item
S. Weinberg, {\it Gravitation and Cosmology} (Wiley, New York, 1972).

\item
G. Parisi, {\it Statistical Field Theory} (Addison-Wesley, Menlo Park,
1988).

\item
D.J. Amit, {\it Field Theory, the Renormalization Group, and Critical
Phenomena}, Second Edition (World Scientific, Singapore, 1984).

\item
J.W. Negele and H. Orland, {\it Quantum Many-Particle Systems}
(Addison-Wesley, Menlo Park, 1988).

\item
J. Zinn-Justin, {\it Quantum Field Theory and Critical Phenomena}, Second
Edition (Clarendon, Oxford, 1993).

\item
F.A. Berezin, {\it The Method of Second Quantization} (Academic Press,
New York, 1966).

\item
J. Glimm and A. Jaffe, {\it Quantum Physics, A Functional Integral Point of
View}, Second Edition (Springer-Verlag, Berlin, 1987).

\item
C. Itzykson and J.-B. Zuber, {\it Quantum Field Theory} (McGraw-Hill, New
York, 1980).

\item
S. Weinberg, {\it The Quantum Theory of Fields} (Cambridge University
Press, Cambridge, 1995).

\item
B. Sakita, {\it Quantum Theory of Many-Variable Systems and Fields}
(World Scientific, Singapore, 1985).

\item
G. Roepstorff, {\it Path Integral Approach to Quantum Physics} 
(Springer-Verlag, Berlin, 1994).

\item
R.J. Rivers, {\it Path Integral Methods in Quantum Field Theory}
(Cambridge University Press, Cambridge, 1987).

\item
S.W. Hawking, {\it Hawking on the Big Bang and Black Holes} (World
Scientific, Singapore, 1993).

\item
{\it Euclidean Quantum Gravity}, edited by G.W. Gibbons and S.W. Hawking
(World Scientific, Singapore, 1993).

\item
See, e.g., M.S. Swanson, {\it Path Integrals and Quantum Processes}
(Academic, Boston, 1992), p. 226.

\item
P. Ramond, {\it Field Theory: A Modern Primer}, Second Edition
(Addison-Wesley, Menlo Park, 1989).

\item
N.D. Birrell and P.C.W. Davies, {\it Quantum Fields in Curved Space}
(Cambridge University Press, Cambridge, 1982).

\item
S. Coleman, {\it Aspects of Symmetry} (Cambridge University Press,
Cambridge, 1985).

\item
T. Appelquist, A. Chodos, and P.G.O. Freund, {\it Modern Kaluza-Klein
Theories} (Addison-Wesley, Menlo Park, 1987).

\item
{\it Physics in Higher Dimensions}, Volume 2, edited by T. Piran and S.
Weinberg (World Scientific, Singapore, 1986).

\item
M.J. Duff, B.E.W. Nilsson, and C.N. Pope, {\it Phys. Rep.} {\bf 130}, 1
(1986).

\item
S. Weinberg, in Ref. 33, p. 359.

\item
M. Nakahara, {\it Geometry, Topology, and Physics} (Adam Hilger, Bristol,
1990).

\item
M. Monastyrsky, {\it Topology of Gauge Fields and Condensed Matter}
(Plenum, New York, 1993).

\item
Y. Choquet-Bruhat, C. DeWitt-Morette, and M. Dillard-Bleick,
{\it Analysis, Manifolds, and Physics} (North-Holland, Amsterdam, 1982).

\item
T. Eguchi, P.B. Gilkey, and A.J. Hanson, {\it Phys. Rep.} {\bf 66}, 213 (1980).

\item
F. Dowker, J.P. Gauntlett, G.W. Gibbons, and G.T. Horowitz, {\it Phys. Rev.} D
{\bf 53}, 7115 (1996).

\item
R. Rajaraman, {\it Solitons and Instantons} (North-Holland, Amsterdam,
1982).

\item
A.S. Schwarz, {\it Quantum Field Theory and Topology} (Springer-Verlag,
Berlin, 1993).

\item
A.M. Polyakov, {\it Gauge Fields and Strings} (Harwood, Chur, 1987).

\item
{\it Instantons in Gauge Theories}, edited by M. Shifman (World
Scientific, Singapore, 1994).

\item
M. G\"{o}ckeler and T. Sch\"{u}cker, {\it Differential
Geometry, Gauge Theories, and Gravity} (Cambridge University Press, Cambridge,
1987).

\item
A. Vilenkin and E.P.S. Shellard, {\it Cosmic Strings and Other Topological
Defects} (Cambridge University Press, Cambridge, 1994).

\item
M.J. Duff, R.R. Khuri, and J.X. Lu, {\it Phys. Rep.} {\bf 259}, 213 (1995).

\item
G.G. Ross, {\it Grand Unified Theories} (Benjamin, Menlo Park, 1984).

\item
S. Weinberg, {\it Rev. Mod. Phys.} {\bf 61}, 1 (1989).

\item
R.M. Wald, {\it General Relativity} (University of Chicago Press, Chicago,
1984).

\item
I.R. Kenyon, {\it General Relativity} (Oxford University Press, Oxford,
1990).

\item
P.J.E. Peebles, {\it Principles of Physical Cosmology} (Princeton
University Press, Princeton, 1993).

\end{enumerate}

\end{document}